\documentstyle[12pt,epsfig]{article}
\begin{document}
\baselineskip 20pt
\begin{center}
\baselineskip=22pt \centerline{\large\bf $B\rightarrow K\pi$
Decays with $1/m_b$ Corrections in $QCD$ Factorization} \hoffset
0.2cm \vspace{1cm} \centerline{Tao Huang
\footnote{Email:huangtao@mail.ihep.ac.cn}, Lin Li$^{b}$, Xue-Qian
Li$^{a,\,\,c}$ Zuo-Hong Li$^{a,\,\,d}$ and Xiang-Yao Wu$^{c}$}
\noindent{\footnotesize a. \textit{CCAST (The World Laboratory),
P.O. Box 8730, Beijing 100080, China}}\\
{\footnotesize b. \textit{Institute of High Energy Physics,
P.O.Box 918(4), Beijing 100039, China}}
\footnote{Mailing address}\\
{\footnotesize c. \textit{Department of Physics, Nankai
University, Tianjin 300071, China}} \\
{\footnotesize d. \textit{Department of Physics, Yantai
University, Yantai 264005, China}}
\end{center}

\begin {minipage}{140mm}
\begin{center}
{\bf Abstract}
\end{center}

A careful investigation of $1/m_b$ power suppressed effects is
crucial for enhancing predictive power of the $QCD$ factorization
approach to charmless B decays. It is instructive to
systematically investigate the $1/m_b$ effects from soft and hard
gluon exchanges, in addition to annihilation topology in the
charmless B decays. In this work we try to give a systematic
discussion on impact of such soft and hard exchanges on the
penguin-dominated $B\to K\pi$ decays within the framework of the
light-cone QCD sum rules (LCSR). For the weak phase $\gamma
(=\textmd{Im}V_{ub}^*)$ ranging from $40^{\circ}$ to $80^{\circ}$,
we find that because of the annihilation and soft and hard
exchanges a numerical increase of $(20-30)\%$ is expected in the
branching ratios up to ${\cal O}(1/m_b^2)$ corrections; the
resultant soft and hard corrections are less important than the
annihilation contributions and amount only to a level of $10\%$.
Possible sources of uncertainties are discussed in some details.
\end {minipage}

\vspace*{2cm} {\bf PACS number(s): 12.38.Lg, 11.55.Hx, 13.20.He}

\vspace{2cm}

\newpage \baselineskip 20pt
\hoffset -0.5cm
\begin{center}
\bf{I. INTRODUCTION}
\end{center}

A considerable progress has been achieved in theoretical study of
$B$ decays since the naive factorization ansatz \cite{BSW} was
proposed for the non-leptonic decays of heavy-mesons. With the QCD
background one makes all effort to approach the physics of B
decays, by developing various theoretical frameworks such as large
energy effective theory (LEET)\cite{LEET}, three-scale
perturbative QCD factorization theorem \cite{HN}, generalized
factorization \cite{Ali}, QCD factorization \cite{BN} and
soft-collinear effective theory (SCET) \cite{SCET1, SCET2, SCET3}.
Especially, more attentions are paid to the QCD factorization and
SCET. Both of them are formulated from the first principle of QCD
and provide a rigorous theoretical basis to the factorization for
a large class of non-leptonic B decays. In comparison, the QCD
factorization approach turns out to be much simpler in structure
and more intuitive in physical picture, while SCET possesses
stronger predictive powers and provides a more complete
theoretical framework, in which, for instance, an elegant proof of
the factorization for B decays into two light mesons and into
$D\pi$ is given in Ref.\cite{SCET2} and some relations can be
established among different decay modes \cite{SCET3}.

The QCD factorization formula has been proved to be a substantial
improvement of the naive factorization assumption. It allows us to
compute radiative corrections to factorizable amplitudes to all
orders in the heavy quark limit $m_b\to\infty$, meanwhile leaving
the $1/m_b$ power-suppressed corrections to be estimated with help
of concrete models. The reasoning behind this theory is that as
one works in the heavy quark limit, the soft gluons with momentum
of order $\Lambda_{QCD}$ decouple and so the interaction kernel
responsible for the transitions can be calculated with the
perturbative QCD (PQCD) in the case of the B decays into two light
mesons. To be specific, the hadronic matrix elements for, say, $B
\rightarrow \pi \pi$ can be expressed as,
\begin{equation}
\langle\pi \pi|O_{i}|B\rangle=\langle\pi|j_{1}|B\rangle \langle
\pi |j_{2}|0\rangle[1+\sum r_{n}
\alpha_{s}^n+O(\frac{\Lambda_{QCD}}{m_{b}})],
\end{equation}
where $O_{i}$ are the concerned local four-quark operators in the
weak effective Hamiltonian, $j_{1,2}$ are the bilinear quark
currents and the other two terms correspond to the perturbative
corrections and non-perturbative contributions respectively.
Nevertheless, an existing problem in the approach is that while
the power corrections in $\alpha_s$ can be calculated in a
systematical way, the $1/ m_b$ power-suppressed effects cannot.
Thus an accurate theoretical prediction on nonleptonic $B$ decays
still is a challenge.

In the QCD factorization, there are a variety of sources of
power-suppressed effects. Among them the annihilation topology,
soft and hard exchanges and final state interaction $(FSI)$
effects etc. are the main ones. The recent work of Mantry, Pirjol
and Stewart \cite{SCET3} indicates that there exists an additional
source of power suppressed contributions. It is obvious that
including all the power-suppressed effects is so far almost
impossible in any practical computation. But it is plausible to
make an order of magnitude estimate of the overall
power-suppressed effects by calculating the effects due to
annihilation topology and soft and hard-gluon exchanges. To this
end, a better understanding or at least a reliable estimation of
the order of magnitude for both effects is crucial to enhance the
predictive power of the QCD factorization. Based on PQCD
\cite{BN1}, contributions of annihilation  to $ B\to \pi\pi, K\pi$
have been estimated, showing a subleading behavior in $1/m_b$.
Soft effects can be understood as processes where a background
field gluon is exchanged between, for instance, an emitted meson
and the other meson which picks up the spectator quark in the case
of emission topology. Therefore, they can be viewed as a
higher-twist effect. Power suppressed hard effects originate from
hard gluon absorption by a spectator quark, which is only relevant
to the penguin contractions of effective operators and the
chromomagnetic dipole operator. There have already been some
earlier attempts \cite{Soft} to understand such soft effects, and
recently a systematic discussion is given by Khodjamiriani
\cite{Kh1}. In Ref.\cite{Kh1} the author suggests using the QCD
light-cone sum rule(LCSR), which is originally developed in
Ref.\cite{LCSR}, to evaluate the non-factorizable corrections to B
decays into two light mesons. Using the generalized LCSR technique
\cite{Kh1, Wu, Kh2} non-factorizable soft corrections to $B\to
\pi\pi$ have been discussed in details. It is found that despite
its numerical smallness, the ${\cal O}(1/m_b)$ soft effect is at
the same level as the corresponding ${\cal O}(\alpha_s)$
corrections to the factorizable amplitudes. This implies that such
soft effects are indispensable for an accurate evaluation of
nonfactorizable contributions to charmless B decays.

Recently, the CLEO-II and-III, Belle and BaBar Collaborations
reported their data on the branching ratios of the $B\rightarrow K
\pi$ decays \cite{Data}. The experimental averages are\cite{Dataa}
\begin{eqnarray}
&&{\cal B}(B^-\rightarrow \bar{K}^0 \pi^-)=(21.8 \pm 1.4
)\times10^{-6}, \nonumber \\&& {\cal B}(\bar{B}^0\rightarrow K^-
\pi^+)=(18.2 \pm 0.8)\times10^{-6}, \nonumber
\\&& {\cal B}(B^-\rightarrow K^-
\pi^0)=(12.5 \pm 1.0)\times10^{-6}, \nonumber
\\&& {\cal B}(\bar{B}^0\rightarrow \bar{K}^0
\pi^0)=(11.7 \pm 1.4)\times10^{-6}.
\end{eqnarray}
Contrast to $B\to \pi\pi$ decays, these decay modes are
penguin-dominated, for the tree contributions are CKM suppressed.
As well known, all new physics effects can only manifest
themselves via loops, so that these processes deserve a detailed
investigation in search for new physics. In this work we will make
an  evaluation of the power-suppressed effects in the
$B\rightarrow K \pi$ decays by investigating all the possible
soft-gluon effects, as well as the power-suppressed hard effects
associated with the penguin topology and chromomagnetic dipole
operator in the framework of the LCSR.

Our paper is organized as follows. In the following section, we
present a systematic LCSR analysis of the soft contributions to
$B\rightarrow K\pi$. The calculation is performed to ${\cal
O}(1/m_b)$. In particular, the soft effects stemming from emission
topology are investigated at some length. Also, the special role
of the penguin topology and chromomagnetic dipole operator is
discussed. They contribute, in addition to a soft effect, a
$1/m_b$-suppressed hard effect which is absent in the QCD
factorization. We estimate both of them with help of the LCSR
results for $B\to \pi\pi$ \cite{Kh2}. In section III, we give the
formula for calculating the branching ratios at subleading orders
in $\alpha_s$ and $1/m_b$, including all the estimated
power-suppressed corrections. In Sec.IV, we present our numerical
discussions along with all the necessary input parameters, and a
somewhat detailed analysis on the main sources of uncertainties,
which influence our numerical results more or less. The last
section is devoted to a brief summary.

\begin{center}
\bf{II. LCSR FOR SOFT-GLUON EFFECTS IN $B\rightarrow K\pi$}
\end{center}

We begin with the weak effective Hamiltonian $H_{eff}$ for the
$\triangle B=1$ transition as \cite{Buras}
\begin{equation}
{\cal
H}_{eff}=\frac{G_{F}}{\sqrt{2}}\sum_{p=u,c}\lambda_p(C_{1}(\mu)O_{1}(\mu)^p+
C_{2}(\mu)O_{2}(\mu)^p
+\sum_{i=3}^{10}C_{i}(\mu)O_{i}(\mu)+C_{8g}(\mu)O_{8g}(\mu))+h.c.,
\end{equation}
where $\lambda_p=V_{pb}V_{pq}^*(q=d,s)$, $C_{i}(\mu)$ and
$C_{8g}(\mu)$ denote the Wilson coefficients, $O_{1,\,2}$, $O_{i}
(i=3-6)$ $(O_{i} (i=7-10))$ and $O_{8g}$ are the tree, QCD
(electroweak) penguin and chromomagnetic dipole operators,
respectively. For a completeness, we list the relevant effective
operators below
\begin{eqnarray}
&&O_{1}^p=(\bar{p}b)_{V-A}(\bar{s}p)_{V-A}, \nonumber  \\&&
O_{2}^p=(\bar{p}_{\alpha}b_{\beta})_{V-A}
(\bar{s}_{\beta}p_{\alpha})_{V-A}, \nonumber  \\&&
O_{3}=(\bar{s}b)_{V-A}\sum_{q}(\bar{q}q)_{V-A}, \nonumber  \\&&
O_{4}=(\bar{s}_{\alpha}b_{\beta})_{V-A}
\sum_{q}(\bar{q}_{\beta}q_{\alpha})_{V-A}, \nonumber \\&&
O_{5}=(\bar{s}b)_{V-A}\sum_{q}(\bar{q}q)_{V+A}, \nonumber
\\&&
O_{6}=(\bar{s}_{\alpha}b_{\beta})_{V-A}
\sum_{q}(\bar{q}_{\beta}q_{\alpha})_{V+A}, \nonumber \\&&
O_{7}=(\bar{s}b)_{V-A}\sum_{q} \frac{3}{2} e_{q}(\bar{q}q)_{V+A},
\nonumber\\&& O_{8}=(\bar{s}_{\alpha}b_{\beta})_{V-A}
\sum_{q}\frac{3}{2}e_{q}(\bar{q}_{\beta}q_{\alpha})_{V+A},
\nonumber\\&& O_{9}=(\bar{s}b)_{V-A}\sum_{q}
\frac{3}{2}e_{q}(\bar{q}q)_{V-A}, \nonumber \\&&
O_{10}=(\bar{s}_{\alpha}b_{\beta})_{V-A}
\sum_{q}\frac{3}{2}e_{q}(\bar{q}_{\beta}q_{\alpha})_{V-A},
\nonumber \\&&
O_{8g}=\frac{-g_s}{8\pi^2}m_b\bar{s}\sigma^{\mu\nu}(1+\gamma_5)G_{\mu\nu}b,
\end{eqnarray}
where $\alpha$ and $\beta$ are the color indices, $q$ runs over
$u, d, s, c$ and $b$, $G_{\mu\nu}\equiv
G_{\mu\nu}^a\frac{\lambda^a}{2}$ is the gluon field strength.

To estimate the soft-exchange effects from emission topology in
the $B\to K\pi$ decays, one would decompose the relevant 4-quark
effective operators into a color singlet part and a color octet
one. Then the soft corrections to the matrix elements of the color
octet operators can be estimated by studying the interactions
between an emitted quark-antiquark pair and a background field
gluon. In the first place, we concentrate ourselves on the case of
$\bar{B}^0\rightarrow K^{-}\pi^{+}$ and calculate such effects in
terms of the LCSR method. The operators, which may induce
soft-exchange effects, contain the tree operator $O_2$, QCD
penguin operators $O_{3,5}$, and EW penguin operators $O_{7,9}$.
The color-octet operators which we encounter in this work, are of
two types of structure: $(V-A)(V-A)$ and $(S+P)(S-P)$. The former
case has been taken into account for $B\to \pi \pi$ in Ref.
\cite{Kh1, Wu}. Here we would like to provide a detailed
derivation of the soft contributions arising from this type of
operators in the $\bar{B}^0\rightarrow K^{-}\pi^{+}$ case. To be
definite, we are going to calculate the soft correction to the
matrix element of the operator
\begin{eqnarray}
\overline{O}=(\overline{u}\frac{\lambda^a}{2}b)_{V-A}
(\overline{s}\frac{\lambda^a}{2}u)_{V-A}.
\end{eqnarray}
Given that the $K^-$ meson is produced as an emitted hadron, the
vacuum-pion correlation function, as the beginning point of the
LCSR calculation, is written as
\begin{equation}
F_{\alpha}(p,q,k)=-\int d^4xe^{-i(p-q)\cdot x}\int
d^4ye^{i(p-k)\cdot y}\langle0|
T\{j_{\alpha5}^{(K)}(y)\overline{O}(0)j_{5}^{B}(x)\}|\pi(q)\rangle,
\end{equation}
where $j_{\alpha5}^{(K)}=\overline{u}\gamma_{\alpha}\gamma_{5}s$
and $j_{5}^{(B)}=m_{b}i\overline{b}\gamma_{5}d$ are the
interpolating fields for $K^{-}$ and $\bar{B}^0$ mesons
respectively. Then we decompose the correlation function (7) with
respect to the independent momenta into four invariant pieces:
\begin{equation}
F_{\alpha}(p,q,k)=(p-k)_{\alpha}F + q_{\alpha}F_{1}+
k_{\alpha}F_{2}+ \epsilon_{\alpha\beta\lambda\rho}q^{\beta}
p^{\lambda}k^{\rho}F_{3}.
\end{equation}
Here an unphysical 4-momentum $k\neq0$ is introduced as an
auxiliary external momentum in the weak operator vertex. Thus the
total momentum of the final state becomes $P=p-k-q$, which is
independent of the momentum $p-q$ in the B channel. The advantage
of introducing $k$ is to help avoiding artificial ambiguities in
the dispersion relation for the B-meson channel. Of course, the
unphysical $k$ has to vanish automatically in the physical matrix
elements. This can be guaranteed, as will be seen, by choosing
kinematical regions in such a way that we let $k^2=0$.

Saturating the correlator (6) with a complete set of intermediate
states of the $K$ quantum numbers and utilizing the definition of
the $K$ meson decay constant
$<0\mid\bar{u}\gamma_{\alpha}\gamma_5s|K^{-}(p-k)>=if_K(p-k)_{\alpha}$,
it follows that only the invariant function $F$ is relevant to our
concern. Explicitly, the resultant hadronic expression for $F$
reads,
\begin{equation}
F_H((p-k)^2, (p-q)^2, P^2)=\frac{if_{K}\Pi_{\pi K}((p-q)^2,
P^2)}{m_K^2-(p-k)^2}+\int_{s_K^0}^{\infty}ds\frac{\rho_h^{(K)}(s,(p-q)^2,
P^2)}{s-(p-k)^2},
\end{equation}
where $s^K_0$ is the threshold parameter and the spectral function
$\rho_h^{(K)}$ stands for the higher state contribution in the K
channel and $\Pi_{\pi K}((p-q)^2, P^2)$ is a correlator,
\begin{equation}
\Pi_{\pi K}((p-q)^2, P^2)=i\int d^4x e^{i(p-q)\cdot x} \langle
K^{-}(p-k)|T \{\overline{O}(0)j_5^{(B)}(x)\}|\pi^-(q)\rangle.
\end{equation}
On the other hand, applying the Operator Product Expansion (OPE)
to Eq.(6) $F$ can be calculated in large space-like regions of
both $p-k$ and $p-q$. In terms of the quark-hadron duality and
then making the Borel transformation $(p-k)^2\longrightarrow M^2$,
we have
\begin{equation}
\Pi_{\pi K}((p-q)^2, P^2)=-\frac{i}{\pi f_K}\int_0^{s_0^K} ds
\textmd{Im} F_{QCD}(s,(p-q)^2, P^2)e^{\frac{m_K^2-s}{M^2}}.
\end{equation}
The correlator $\Pi_{\pi K}((p-q)^2,P^2)$ is applicable for only
large space-like $ P^2$, therefore one needs to make an analytic
continuation of Eq.(10) from a large space-like region $P^2\ll 0$
to a large time-like region $p^2=m_B^2$ for the realistic
$\bar{B}^0\to K^-\pi^+$ decay. We then have
\begin{eqnarray}
\Pi_{\pi K}((p-q)^2, M_B^2)&=&i\int d^4x e^{i(p-q)\cdot x} \langle
K^{-}(p-k)\pi^+(-q)|T \{\overline{O}(0)j_5^{(B)}(x)\}|0\rangle\nonumber\\
&=&-\frac{i}{\pi f_K}\int_0^{s_0^K} ds \textmd{Im}
F_{QCD}(s,(p-q)^2, M_B^2)e^{\frac{m_K^2-s}{M^2}}.
\end{eqnarray}
Inserting further the intermediate states of $\bar{B^0}$ quantum
numbers in the correlator of Eq (11), the hadronic matrix element
$<K^-\pi^+|\overline{O}|B>$ can be extracted in the light of the
standard procedure for the QCD sum rule calculation. Not giving
any technical details, we end up with the following LCSR
expression for the matrix element in question,
\begin{eqnarray}
\langle K^-(p)\pi^+(-q)|\overline{O}|\bar{B}(p-q)
\rangle&=&\frac{-i}{f_{K}f_{B}{m_{B}}^2}\int_{0}^{s_{0}^{K}}dse^
{\frac{m_K^2-s}{M^2}}\int_{m_b^2}^{\bar{R}(s,m_b^2,m_B^2,s_0^B)}
ds^{\prime}
\rho_{QCD}(s,s',m_B^2)\nonumber\\
&\times& e^{\frac{m_B^2-s^{\prime}}{{M^{\prime}}^2}},
\end{eqnarray}
where $f_{B}$ is the B decay constant, $s_0^B$ and ${M^\prime}^2 $
are the threshold and Borel parameters in the B-channel and the
QCD double spectral density
$\rho_{QCD}(s,s',m_B^2)=1/\pi^2\textmd{Im}_{s}\textmd{Im}_{s^{\prime}}F_{QCD}(s,s^{\prime},m_B^2).$
We will use the notation $A_1$ to denote such matrix element from
now on.

Now we expand the correlation function (6) near the light-cone
$x^2 \sim y^2 \sim (x-y)^2 \sim 0$ in order to get
$\rho_{QCD}(s,s',m_B^2)$. The kinematical regions we choose are
summarized as follows:
\begin{equation}
q^2=0,p^2=m_K^2, k^2=0,
\end{equation}
and
\begin{equation}
(p-k)^2\sim (p-q)^2\sim P^2\to -\infty,
\end{equation}
where the LCSR calculation is efficient and self-consistent.
Moreover, we set all the light quark masses to be zero and neglect
all the terms proportional to order of $1/m_b^2$ which emerge in
the calculations. As has been mentioned, the soft effect is due to
the interactions between emitted quark-antiquark pairs and a
background field gluon, thus the underlying quark propagator would
receive a correction term $\cite{Gp}$:
\begin{equation}
\overline{S}(x,0)=\frac{g_s\Gamma(n/2-1)}{16\pi^2(-x^2)^{n/2-1}}
\int\limits_{0}^{1}dv
\{(1-v)x\!\!\!/\sigma_{\mu\nu}G^{\mu\nu}(vx)+v\sigma_{\mu\nu}G^{\mu\nu}(vx)x\!\!\!/\}
\end{equation}
where $n$ is the space-time dimension.

Using Eq.(15) and parameterizing the nonperturbative QCD effects
with the three particle distribution functions of the pion, a
straightforward calculation leads to the following light-cone QCD
result
\begin{equation}
F_{QCD}=F_{tw3}+ F_{tw4},
\end{equation}
with the twist-3 contribution
\begin{eqnarray}
F_{tw3}&=&\frac{m_b f_{3{\pi}}}{4 {\pi}^2}\int_{0}^{1} dv \int
D\alpha_{i}\frac{\varphi_{3
\pi}(\alpha_{i})}{(m_b^2-(p-q)^2(1-\alpha_{1})) (-P^2v
\alpha_{3}-(p-k)^2(1-v \alpha_{3}))} \nonumber \\
&\times&[(2-v)(q \cdot k)+2(1-v)q \cdot(p-k)]q \cdot(p-k),
\end{eqnarray}
and the twist-4 contribution
\begin{eqnarray}
F_{tw4}&=&-\frac{m_b^2f_{\pi}}{{\pi}^2} \left(\int_{0}^{1}dv \int
D\alpha_{i}\widetilde{\varphi}_{\perp}(\alpha_{i})\frac
{1}{2[m_b^2-(p-(1-\alpha_{1})q )^2]}\frac
{(2v-3)(p-k)\cdot q}{(p-k-v\alpha_{3}q)^2}\right. \nonumber \\
&-&\left.\int_{0}^{1}dv \int d\alpha_{1} d\alpha_{3}
\Psi_{1}(\alpha_{1},\alpha_{3})\frac {1}{[m_b^2-(p-(1-
\alpha_{1})q )^2]^2}\frac{(p\cdot q-vq\cdot k)(p-k)\cdot q}
{(p-k-v\alpha_{3}q)^2}\right.\nonumber \\
&+&\left.\int_{0}^{1}dv \int d\alpha_{3} \Psi_{2}
(\alpha_{3})\frac {1}{[m_b^2-(p-\alpha_{3}q)^2]^2}\frac{(p\cdot
q-vq\cdot k)(p-k)\cdot q}
{(p-k-v\alpha_{3}q)^2}\right.\nonumber \\
&-&\left.\int_{0}^{1}dv v^2\int d\alpha_{3} \Psi_{2}
(\alpha_{3})\frac {1}{p\cdot
q[m_b^2-(p-\alpha_{3}q)^2]}\frac{[(p-k)\cdot q]^3}
{(p-k-v\alpha_{3}q)^4}\right.\nonumber \\
&+&\left.\int_{0}^{1}dv(v-1)v\int d\alpha_{3} \Psi_{2}(\alpha_{3})
\frac {1}{m_b^2-(p-\alpha_{3}q)^2}\frac{[(p-k)\cdot q]^2}
{(p-k-v\alpha_{3}q)^4}\right).
\end{eqnarray}
In Eqs.(17) and (18), $D\alpha_i\equiv d\alpha_1 d\alpha_2
d\alpha_3 \delta(1-\alpha_1-\alpha_2 -\alpha_3)$; $f_{\pi}$ is the
decay constant of the pion, $f_{3\pi}$ indicates a nonperturbative
parameter defined by the matrix element
$<0|\bar{d}\sigma_{\mu\nu}\gamma_5g_sG_{\alpha\beta}u|\pi^+>$;
$\varphi_{3\pi}(\alpha_i)$ is a pionic twist-3 distribution
function, whereas $\widetilde{\varphi}_{\perp}(\alpha_i)$ is of
twist-4 and it, together with another twist-4 distribution
amplitude $\widetilde{\varphi}_{\parallel}(\alpha_{i})$, defines
the functions $\Psi_{1,2}$ as
\begin{eqnarray}
&&\Psi_{1}(u,v)=\int_0^u d\eta \left.(\widetilde{\varphi}_{\perp}
(\eta,1-\eta-v,v)+\widetilde{\varphi}_{\parallel}(\eta,1-\eta-v,v)\right.), \nonumber\\
&&\Psi_{2}(u)=\int_0^u d\eta \int_0^{1-\eta}d\xi
\left.(\widetilde{\varphi}_{\perp}(\xi, 1-\xi-\eta,
\eta)+\widetilde{\varphi}_{\parallel}(\xi,1-\xi-\eta,\eta)\right.).
\end{eqnarray}
Readers are advised to refer to Ref.\cite{RK} for the definitions
of various wavefunctions involved here and hereafter.

A simple manipulation can make Eq.(16) expressed in a preferred
form of dispersion integral, from which extraction of
$\rho^{QCD}(s,s',m_B^2)$ is straightforward. Substituting the
yielded $\rho^{QCD}(s,s',m_B^2)$ into Eq.(12), we derive the final
LCSR results for $A_1$ as
\begin{eqnarray}
A_{1}&=& im_{B}^2 \left(\frac{1}{4{\pi}^2 f_{K}}
\int_{0}^{s_{0}^{K}}dse^{\frac{m_K^2-s}{M^2}}\right)\left(\frac{m_b^2}{2f_{B}
m_{B}^4}\int_{u_0^B}^{1} \frac{du}{u}
e^{\frac{m_{B}^2}{{M^{\prime}}^2}-\frac{m_b^2}{u{M^{\prime}}^2}}\right.
\nonumber \\&&
\left.\times\left[\frac{m_bf_{3\pi}}{u}\int_{0}^{u}\frac{dv}{v}\varphi_{3\pi}
(1-u,u-v,v)+f_{\pi}\int_{0}^{u}\frac{dv}{v} \Large[3
\widetilde{\varphi}_{\perp} (1-u,u-v,v)\right.\right.\nonumber
\\&&
-\left.\left.\left(\frac{m_b^2}{u{M^{\prime}}^2}-1\right)\frac{\Psi_{1}(1-u,v)}
{u}\Large]+f_{\pi}\left(\frac{m_b^2}{u{M^{\prime}}^2}-2\right)\frac{\Psi_{2}(u)}{u^2}\right]\right),
\end{eqnarray}
where $u_0^B=m_b^2/s_0^B$. It is noted that for the twist-3 part
we obtain the same result as that in $B\to \pi \pi$ case
\cite{Kh1}, whereas the obtained twist-4 parts are not quite the
same (in contrast with the corresponding term
$(\frac{m_b^2}{u{M^{\prime}}^2}-
\frac{s}{{M^{\prime}}^2}-1)\frac{\Phi_2(u)}{u^2}$ in Ref.
\cite{Kh1}, our result is
$(\frac{m_b^2}{u{M^{\prime}}^2}-2)\frac{\Phi_2(u)}{u^2}$).
Numerically, however, the two forms result in close numbers.

Applying the same procedure to the $(S+P)(S-P)$ operators, one can
notice that they do not result in any soft contributions to the
amplitudes at all.

Now let us turn to a discussion on the other three decay modes.
The case of $B^-\to \bar{K^0} \pi^-$ is simple, where only the $
(V-A)(V-A)$ operators $O_{3,9}$ are concerned so that the LCSR
result (20) may apply directly. The situations for the $B^-\to
K^-\pi^0$ and $\bar{B^0}\to \bar{K^0}\pi^0$ decays are a bit
complicated, in which both $K$ mesons may either be an emitted
hadron or include a spectator quark (antiquark). It is obvious
that the LCSR result (20) holds for the emission case where only
the  $(V-A)(V-A)$ operators contribute to the decay amplitudes
(note that there is an additional factor of $1/\sqrt{2}$
$(-1/\sqrt{2})$ for the operator
$(\bar{u}\frac{\lambda^a}{2}b)_{V-A}(\bar{s}\frac{\lambda^a}{2}u)_{V-A}$
$((\bar{d}\frac{\lambda^a}{2}b)_{V-A}(\bar{s}\frac{\lambda^a}{2}d)_{V-A})$.
For the latter case, however, we have to modify the correlation
function (6) with a necessary replacement, and besides the
$(V-A)(V-A)$ and $(S+P)(S-P)$ operators we have  to deal with the
operators of $(V-A)(V+A)$ structure. Omitting the concrete
derivations to save space, we  only present a simple summary of
our results. The operator
$(\bar{s}\frac{\lambda^a}{2}b)_{V-A}(\bar{u}\frac{\lambda^a}{2}u)_{V-A}$
$((\bar{s}\frac{\lambda^a}{2}b)_{V-A}(\bar{d}\frac{\lambda^a}{2}d)_{V-A})$
provides the relevant matrix elements associated with the soft
contribution $1/\sqrt{2}A_{2}$ $(-1/\sqrt{2}A_{2})$,
\begin{eqnarray}
A_{2}&=&im_{B}^2\left(\frac{1}{4{\pi}^2 f_{\pi}}
\int_{0}^{s_{0}^{\pi}}dse^{-\frac{s}{M^2}}\right)
\left(\frac{m_b^2}{2f_{B} m_{B}^4}\int_{u_0^B}^{1} \frac{du}{u}
e^{\frac{m_{B}^2}{{M^{\prime}}^2}-\frac{m_b^2}{u{M^{\prime}}^2}}
\right.\nonumber \\&&
\left.\times\left[\frac{m_bf_{3K}}{u}\int_{0}^{u}\frac{dv}{v}\varphi_{3k}
(1-u,u-v,v)+f_{K}\int_{0}^{u}\frac{dv}{v}[3
\widetilde{\varphi}_{\perp} (1-u,u-v,v)\right.\right. \nonumber
\\&&
\left.\left.-(\frac{m_b^2}{u{M^{\prime}}^2}-1)\frac{\Psi_{1}(1-u,v)}
{u}]+f_{K}\left(\frac{m_c^2}{u{M^{\prime}}^2}-2\right)\frac{\Psi_{2}(u)}{u^2}\right]\right).
\end{eqnarray}
The $(S+P)(S-P)$ operators make a vanishing contribution, as in
the $K$ emission case. It is interesting to notice that the
$(V-A)(V+A)$ operators have an equal matrix element to those of
the corresponding $(V-A)(V-A)$ operators.

Expanding the LCSR results for $A_1$ and $A_2$ in $1/m_b$ and then
comparing them with the corresponding factorizable amplitudes, we
observe that the soft effects of the emission topology are
typically of order $1/m_b$.

Finally, the soft contributions of the emission topology to the
$B\to K\pi$ decay amplitudes can be parametrized in terms of the
resultant $A_1$ and $A_2$ as the following,
\begin{eqnarray}
M_{s}^{(O_i)}(B^-\rightarrow
\bar{K}^0\pi^{-})=-\frac{G_F}{\sqrt{2}}
V_{tb}V_{ts}^*(2C_3-C_9)A_1,\nonumber
\end{eqnarray}
\begin{eqnarray}
M_{s}^{(O_i)}(\bar{B}^0\rightarrow K^-\pi^{+})=\sqrt{2}G_F(
V_{ub}V_{us}^*C_2A_1-V_{tb}V_{ts}^*(C_3+C_9))A_1,\nonumber
\end{eqnarray}
\begin{eqnarray}
M_{s}^{(O_i)}(B^-\rightarrow K^-\pi^{0})&=&G_F\{
V_{ub}V_{us}^*(C_1A_2+C_2A_1)-V_{tb}V_{ts}^*[(C_3+C_9)A_1 \nonumber\\
&&-\frac{3}{2}(C_8A_2+C_{10})A_2]\},
\end{eqnarray}
\begin{eqnarray}
M_{s}^{(O_i)}(\bar{B}^0\rightarrow \bar{K}^0\pi^{0})&=&G_F\{
V_{ub}V_{us}^*C_1A_2+V_{tb}V_{ts}^*[(C_3-\frac{1}{2}C_9)A_1\nonumber\\
&&-\frac{3}{2}(C_8+C_{10})A_2]\}.
\end{eqnarray}
which evidently respect the isospin symmetry.

Besides the soft contributions due to the emission topology, the
two-body B decays, generally speaking, receive the
power-suppressed corrections from the chromomagnetic diploe
operator $O_{8g}$ and penguin topology \cite{Kh2}. The relevant
contributions contain a soft and a hard parts. The former is owing
to a soft gluon which is emitted off either from the $O_{8g}$
vertex or from a quark loop of the penguin contraction and finally
immerses into the meson which absorbs a spectator quark. The
latter is of two different origins. One is the hard gluon exchange
between the $O_{8g}$ or penguin vertices and a spectator
quark(antiquark). Another is related to the factorizable quark
condensate contributions with QCD radiative correction being
involved. The LCSR approach also is suitable for a quantitative
study on these corrections. In fact, a detailed discussion has
been made on their influences on $B\to \pi\pi$ in the LCSR
framework \cite{Kh2}. The behaviors of such contributions in the
heavy quark limit $m_b\to \infty$ comply with the following power
counting: (1) The penguin diagrams make no contributions up to
$1/m_b^2$ order. (2) The $O_{8g}$ operator supplies a hard effect
of ${\cal O}(1/m_b)$ and a soft effect of ${\cal O}(1/m_b^2)$.
Despite being formally suppressed by ${\cal O}(1/m_b^2)$ compared
with the leading-order factorizable amplitudes, the soft
correction from $O_{8g}$ is free of $\alpha_s$ suppression and
numerically comparable with the ${\cal O}(\alpha_s)$ hard part.
(3) The part with quark condensate has a chiral enhancement factor
$r_{\chi}^{\pi}$ via the PCAC relation and it suffers formally
from ${\cal O}(\alpha_s)$ and ${\cal O}(1/m_b)$ double suppression
but numerically turns out to be a large effect. However, it has a
counterpart in the $QCD$ factorization and thus is not included in
our calculation to avoid double counting.

It is reasonable to assume that the same power counting holds for
$B\to K\pi$. Furthermore, since no strange quark appears as a
spectator in the concerned case, the emitted soft gluons  can only
combine with a quark pair to form a three-particle Fock state of
the pion, we may directly use the corresponding LCSR results
\cite{Kh2} by simply replacing the relevant parameters to achieve
an estimate of the soft and hard effects due to the $O_{8g}$
operator and penguin topology in the $B\to K\pi$ case. As a
consequence, the resultant corrections of $O_{8g}$ to the decay
amplitudes can be expressed, at the subleading order in $1/m_b$ as
follows,
\begin{eqnarray}
M_{h+s}^{(O_{8g})}(B^-\rightarrow \bar{K}^0
\pi^-)&=&M_{h+s}^{(O_{8g})}(\bar{B}^0\rightarrow \bar{K}^-
\pi^+)\nonumber\\&=&\sqrt{2}M_{h+s}^{(O_{8g})}(B^-\rightarrow K^-\pi^{0})\nonumber\\
&=&-\sqrt{2}M_{h+s}^{(O_{8g})}(\bar{B}^0\rightarrow
\bar{K}^0\pi^{0})\nonumber\\&=& -\frac{G_F}{\sqrt{2}}
V_{tb}V_{ts}^* C_{8g}(A_h^{(O_{8g})}+A_s^{(O_{8g})}).
\end{eqnarray}
Here $A_{h}^{(O_{8g})}$ and $A_{s}^{(O_{8g})}$ express the hard
and soft parts of the matrix element $<\bar{K}^0
\pi^-|O_{8g}|B^->$, and have the following LCSR results:
\begin{eqnarray}
A_{h}^{(O_{8g})}&=&i\frac{\alpha_sC_F}{2\pi}m_b^2\left(\frac{1}{4\pi^2f_{K}}\int_{0}^{s_{0}^{K}}dse^{(m_K^2-s)/M^2}\right)
\left(\frac{m_b^2f_{\pi}}{2f_{B} m_{B}^2}\int_{u_0^B}^{1}
\frac{du}{u}
e^{m_{B}^2/{M^{\prime}}^2-m_b^2/u{M^{\prime}}^2}\right. \nonumber \\
&&\times\left.\left[\varphi_p(1)\bar{u}\left(1+\frac{3m_b^2}{u
m_B^2}\right)-\frac{\varphi_{\sigma}^{\prime}(1)\bar{u}}{6}\left(\frac{5m_b^2}{u
m_B^2}-1\right)\right]\right),
\end{eqnarray}
\begin{eqnarray}
A_{s}^{(O_{8g})}&=&im_b^2\left(\frac{1}{4\pi^2f_{K}}\int_{0}^{s_{0}^{K}}dse^{(m_K^2-s)/M^2}\right)
\left(\frac{f_{\pi}}{f_{B} m_{B}^2}\int_{u_0^B}^{1} \frac{du}{u}
e^{m_{B}^2/{M^{\prime}}^2-m_b^2/u{M^{\prime}}^2}\right.\nonumber \\
&&\left.\times\left(1+\frac{m_b^2}{u
M_B^2}\right)\left[\varphi_{\perp}
(1-u,0,u)+\widetilde{\varphi}_{\perp} (1-u,0,u)\right]\right),
\end{eqnarray}
where both distribution amplitudes $\varphi_{p}(u)$ and
$\varphi_{\sigma}(u)$ are of twist-3 and are used to describe the
pionic valence Fock state, while $\varphi_{\bot}(u)$ is a twist-4
three-particle wavefunction in analogy to
$\widetilde{\varphi}_{\perp}(\alpha_i)$ and
$\widetilde{\varphi}_{\parallel}(\alpha_{i})$.

\begin{center}
\bf{III. DECAY AMPLITUDES AND BRANCHING RATIOS WITH SUBLEADING
CORRECTIONS}
\end{center}

The LCSR results for the ${\cal O}(1/m_b)$ soft and hard
corrections to the $B\to K\pi$ decay amplitudes may serve as an
order of magnitude estimate of the overall power-suppressed
effects. We add them, together with the annihilation contributions
$M_{a}$, to the $QCD$ factorization results $M_f$, to get a decay
amplitude with the subleading power corrections in both $\alpha_s$
and $1/m_b$. In Ref.\cite{BN1}, the $B\to K\pi$ decay amplitudes
have been computed by including the ${\cal O}(\alpha_s)$
corrections, and the annihilation effects have also been estimated
in PQCD. Their results, which will be used for our upcoming
numerical discussion, are:
\begin{eqnarray}
M_{f}(B^-\rightarrow \bar{K}^0
\pi^-)=\sum_{p=u,c}V_{pb}V_{ps}^*[(a_4^p-\frac{1}{2}a_{10}^p)+
r_{\chi}^{K}(a_6^p-\frac{1}{2}a_8^p)]A_{\pi K},\nonumber
\end{eqnarray}
\begin{eqnarray}
M_{f}(B^-\rightarrow K^-
\pi^0)&=&\frac{1}{\sqrt{2}}([V_{ub}V_{us}^*a_1+\sum_{p=u,c}V_{pb}V_{ps}^*(a_4^p+a_{10}^p)\nonumber\\
&+&\sum_{p=u,c}V_{pb}V_{ps}^*r_{\chi}^K(a_6^p+a_{8}^p)]A_{\pi K}
\nonumber \\&+&[V_{ub}V_{us}^*a_2+
\sum_{p=u,c}V_{pb}V_{ps}^*\frac{3}{2}(-a_7+a_9)]A_{K
\pi}),\nonumber
\end{eqnarray}
\begin{eqnarray}
M_{f}(\bar{B}^0\rightarrow K^-
\pi^+)&=&[V_{ub}V_{us}^*a_1+\sum_{p=u,c}V_{pb}V_{ps}^*(a_4^p+a_{10}^p)\nonumber\\
&+&\sum_{p=u,c}V_{pb}V_{ps}^*r_{\chi}^K(a_6^p+a_{8}^p)]A_{\pi
K},\nonumber
\end{eqnarray}
\begin{eqnarray}
M_{f}(\bar{B}^0\rightarrow \bar{K}^0 \pi^0)&=&M_{f}(B^-\rightarrow
K^- \pi^0)- \frac{1}{\sqrt{2}}M_{f}(B^-\rightarrow \bar{K}^0\pi^-
)\nonumber\\&-&\frac{1}{\sqrt{2}}M_{f}(\bar{B}^0\rightarrow
K^-\pi^+),
\end{eqnarray}
and
\begin{eqnarray}
M_{a}(B^-\rightarrow \bar{K}^0
\pi^-)=[V_{ub}V_{us}^*b_2+(V_{ub}V_{us}^*+V_{cb}V_{cs}^*)(b_3+b_3^{EW})]B_{\pi
K},\nonumber
\end{eqnarray}
\begin{eqnarray}
M_{a}(B^-\rightarrow K^-
\pi^0)=\frac{1}{\sqrt{2}}M_{a}(B^-\rightarrow \bar{K}^0\pi^-
),\nonumber
\end{eqnarray}
\begin{eqnarray}
M_{a}(\bar{B}^0\rightarrow K^-
\pi^+)=(V_{ub}V_{us}^*+V_{cb}V_{cs}^*)(b_3-\frac{1}{2}b_3^{EW})B_{\pi
K},\nonumber
\end{eqnarray}
\begin{eqnarray}
M_{a}(\bar{B}^0\rightarrow \bar{K}^0
\pi^0)=-\frac{1}{\sqrt{2}}M_{a}(\bar{B}^0\rightarrow K^-\pi^+ ).
\end{eqnarray}
In Eq.(26), $a_i$ and $a_i^p$ are the $QCD$ modified effective
coefficients, and the low energy effects for the heavy to light
transitions are included in the parameters $A_{\pi,K}(A_{K,\pi})$
with an obvious dependence on the $B\to \pi (B\to K)$ form factor
$F^{B\to \pi}(m_K^2)$$( F^{B\to K}(m_{\pi}^2))$, the parameter
$r_{\chi}^K$ is a so called chiral enhancement factor which is
related to the running quark masses. As  the parameters existing
in Eq.(27) are concerned, $b_{2}$ and $b_3(b_3^{EW})$ are the
parameters related to the tree operator $O_2^p$ and QCD
(electroweak) penguin operators and embody the QCD dynamics in the
annihilation processes, while $B_{\pi K}$ is closely related to
the decay constants. The explicit definitions of all these
quantities are given in Ref. \cite{BN1} and we do not repeat them
here.

At present, we can write down the decay amplitudes of
$B\rightarrow K\pi$ with the subleading corrections in both
$\alpha_s$ and $1/m_b$,
\begin{eqnarray}
M(B\rightarrow K\pi)&=&M_{f}(B\rightarrow K\pi)+
M_{a}(B\rightarrow
K\pi)\nonumber\\
&+&M_{s}^{(O_i)}(B\rightarrow K\pi)+
M_{h+s}^{(O_{8g})}(B^-\rightarrow K \pi).
\end{eqnarray}
Using the resultant decay amplitudes, it is straightforward to
calculate the branching ratios of the $B\to K\pi$ decays, which
are given by
\begin{equation}
{\cal B}{(B\rightarrow K \pi)} =\frac{{\cal \tau}_B}{8\pi}{\vert
{M(B\rightarrow K \pi)}\vert}^2\frac{\vert{\bf P}\vert}{m_{B}^2},
\end{equation}
where ${\bf P}$ is the c.m momentum of the outgoing mesons in the
center of mass frame of B meson,
\begin{equation}
\vert{\bf P}\vert=\frac{[(m_{B}^2-(m_{K}+m_{\pi})^2)
(m_{B}^2-(m_{K}-m_{\pi})^2)]^{\frac{1}{2}}}{2m_{B}},
\end{equation}
and ${\cal \tau}_B $ is the $B$ lifetimes.

%\newpage
\begin{center}
\bf{IV. NUMERICAL RESULTS AND DISCUSSIONS }
\end{center}

Now we are in a position to make a numerical analysis and then to
see how the subleading effects in $1/m_b$ impact the results of
the $QCD$ factorization. The set of important parameters of the
present concern contains the $QCD$ modified effective
coefficients, quark masses, form factors, decay constants and
distribution amplitudes. For the $QCD$ modified effective
coefficients and current quark masses, we adopt the numerical
values presented in Ref.\cite{BN1}. There have been a number of
model-dependent estimates for the form factors $F^{B\to
\pi,K}(q^2)$ and decay constants $f_B$ in the literature.
Consistently we pick up, as inputs, the sum rule evaluations
\cite{FD}: $F^{B\rightarrow \pi}(0)=0.28\pm 0.06$,
$F^{B\rightarrow K}(0)=0.32\pm 0.05$ and $f_B=180\pm 30\, MeV$
with $m_b=4.7\pm 0.1\, GeV$. The decay constants of $\pi$ and $K$
mesons are taken as $f_{\pi}=132\, MeV$ and $f_{K}=160\, MeV$,
respectively. As for the distribution amplitudes for the pion and
$K$ meson, we don't take into account SU(3) breaking effect for
consistency, and adopt the following forms \cite{RK}:
\begin{eqnarray}
&&\varphi_{\pi, K}(\mu)=6u(1-u) [1+a_2(\mu)C^{3/2}_2(2u-1)],\nonumber \\
&&\varphi_{p}(u,\mu)=1+30\frac{f_{3\pi}}{\mu_{\pi}f_{\pi}}C_2^{1/2}(2u-1)-3\frac{f_{3\pi}}{\mu_{\pi}f_{\pi}\omega_{3\pi}}C_4^{1/2},\nonumber \\
&&\varphi_{\sigma}(u,\mu)=6u(1-u)\left[1+5\frac{f_{3\pi}}{\mu_{\pi}f_{\pi}}(1-\frac{\omega_{3\pi}}{10})
C_2^{3/2}(2u-1)\right],\nonumber \\
&&\varphi_{3\pi}(\alpha_i)=360\alpha_1 \alpha_2 \alpha_3^2\left[1+\frac{\omega_{3\pi}}{2}(7\alpha_3-3)\right],\nonumber \\
&&\varphi_{\perp}(\alpha_i)=30\delta_{\pi}^2(\alpha_1-\alpha_2)\alpha_3^2[\frac{1}{3}+2\epsilon_{\pi}(1-2\alpha_3)],\nonumber \\
&&\tilde{\varphi}_{\parallel}(\alpha_i)=-120\delta_{\pi}^2\alpha_1\alpha_2\alpha_3\left[\frac{1}{3}+\epsilon_{\pi}(1-3\alpha_3)\right],\nonumber \\
&&\tilde{\varphi}_{\perp}(\alpha_i)=30\delta_{\pi}^2(1-\alpha_3)\alpha_3^2\left[\frac{1}{3}+2\epsilon_{\pi}(1-2\alpha_3)\right],
\end{eqnarray}
where the the leading twist-2 distribution amplitudes
$\varphi_{\pi, K}(\mu)$ enter in calculation of the QCD
factorization and annihilation contributions. The various
coefficients in above equation have been determined \cite{RK} at
$\mu_b=\sqrt{m_B^2-m_b^2}\simeq 2.4 \textsl{GeV}$:
$f_{3\pi}=0.0026 \textsl{GeV\,}^2$, $\omega_{3\pi}=-2.18$,
$\mu_{\pi}(=m_{\pi}^2/(m_u+m_d))=2.02 \textsl{GeV}$,
$\delta^2=0.17 \textsl{GeV\,}^2$, $\epsilon=0.36$. The
scale-dependence of them is achievable by use of the
renormalization group equations \cite{Kh2,RK}. The parameters
inherent in the sum rules can be fixed via the two-point sum rules
at: $s_0^{\pi}=0.7 $GeV$^2$ and $M^2=0.5-1.2$ $GeV^2$ \cite{Kh1},
$s_0^{K}=1.2$ $GeV^2$ and $M^2=1.5-3$ $GeV^2$ \cite{Kt};
$s_0^{B}=35\pm 2$ $GeV^2$ and ${M^{\prime}}^2=8-12$ $GeV^2$
\cite{FD}, corresponding to the pion, $K$ and $B$ channels
respectively. The $B$-lifetimes are measured experimentally as
\cite{Data1}: $\tau_{\bar{B}^0}=1.542\times 10^{-12}ps$ and
$\tau_{B^-}=1.674\times 10^{-12}ps$. In addition, to explicitly
investigate the dependence of the resultant branching ratios on
the weak phase $\gamma=\textmd{Im} V_{ub}^*$, it is convenient to
employ the following parametrization for the $CKM$ matrix elements
\cite{Data1}:
\begin{eqnarray}
&&V_{ub}V_{us}^*=(1-\frac{\lambda^2}{2})|V_{cb}|R_b e^{-i\gamma},
\nonumber \\&& V_{cb}V_{cs}^*=(1-\frac{\lambda^2}{2})|V_{cb}|,
\nonumber \\&& V_{tb}V_{ts}^*=-V_{ub}V_{us}^*-V_{cb}V_{cs}^*.
\end{eqnarray}
where $|V_{cb}|=0.0395\pm 0.0017$, $\lambda=|V_{us}|=0.2196$, and
$R_b=\frac{\lambda}{1-\frac{\lambda^2}{2}}|\frac{V_{ub}}{V_{cb}}|$
with $|V_{ub}/V_{cb}|=0.085\pm 0.02$.

Before an explicit numerical analysis is made, we would like to
simply judge the pattern for the rates of $B\to K \pi$ by
analyzing the results of naive factorization. The following
hierarchy among the rates should be expected:
\begin{eqnarray}
\Gamma(B^-\to \bar{K}^0 \pi^{-})\geq \Gamma(\bar{B}^0\to K^-
\pi^{+})>\Gamma(B^- \to K^- \pi^{0})>\Gamma(\bar{B}^0\to \bar{K}^0
\pi^{0}).
\end{eqnarray}
The reasons are the following: (1) The penguin contributions
dominate in these decays, since the tree ones are less important
by the CKM suppression. (2) It is generally desirable that the
branching ratios for $B^-\to \bar{K}^0 \pi^{-}$ and $\bar{B}^0\to
K^- \pi^{+}$ must be very close due to the smallness of the
electroweak penguin effect, the slight difference between them
arises from a destructive interference between the tree and QCD
penguin contributions to $\bar{B}^0\to K^- \pi^{+}$. (3) The
$\pi^0$ wavefunction makes the first two modes in Eq.(33) larger
than the other two, and $\Gamma(B^- \to K^- \pi^{0})$ nearly half
as large as $\Gamma(B\to K \pi^{\pm})$. (4) Despite the fact that
the electroweak penguin has a negligible effect on the first two
modes, it plays a non-negligible role to the last two. With a
moderate electroweak penguin contribution, the interference would
become constructive between the electroweak and QCD penguins in
$B^- \to K^- \pi^{0}$, in contrast to the $\bar{B}^0\to \bar{K}^0
\pi^{0}$ case where the interference is destructive between the
two penguins. Qualitatively the pattern is in agreement with the
experimental observations.

Although the numerical calculation can be done using the inputs
given above, we should note that the uncertainties resulting from
such calculations are difficult to be quantitatively evaluated
because of our poor knowledge about the inputs, especially the
higher-twist distribution amplitudes. However, the yielded results
are adequate to serve as an order of magnitude estimate of the
effects in question.

we present our predictions based on the $QCD$ factorization
formula ${\cal B}^{(f)}(B\to K\pi)$ for each mode, respectively in
Fig.1-4, together with those with the annihilation effects
included ${\cal B}^{(f+a)}(B\to K\pi)$ and the results all the
estimated subleading corrections in $\alpha_s$ and $1/m_b$ ${\cal
B}^{(nl)}(B\to K\pi)$. The yielded ${\cal B}^{(f)}(B\to K\pi)$ and
${\cal B}^{(f+a)}(B\to K\pi)$ are plotted for the four different
modes, respectively in Fig.5 and Fig.6. It should be understood
that only the central values, which are obtained for $\mu=\mu_b$
and $\gamma=60^{\circ} \pm 20^{\circ}$, are shown there. Working
within the QCD factorization framework, one observes from Fig.5:
(1) The yielded branching ratios answer to the pattern
\begin{equation}
{\cal B}^{(f)}(\bar{K}^0\pi^-)>{\cal B}^{(f)}(K^-\pi^+)>{\cal
B}^{(f)}(K^-\pi^0)>{\cal B}^{(f)}(\bar{K}^0 \pi^0).
\end{equation}
(2) The branching ratios of both $ \bar{K}^0 \pi^-$ and $B\to
\bar{K}^0 \pi^0$ are less sensitive to the change of $\gamma$ than
the other two, as expected.

As the $1/m_b$ power-suppressed effects are involved from the
annihilations and soft and hard gluons, the branching rations get
an evident modification as shown in Fig.6.  Letting $\mu$ vary
between $\mu_b/2$ and $2\mu_b$, we find that the $1/m_b$
power-suppressed effects from the annihilations and soft and hard
gluons can enhance the branching ratios by $(20-30)\%$, depending
on the decay modes and $\gamma$.

Now, for taking a closer look at the individual roles of the soft
and hard exchange and annihilation effects in the subleading
contributions in $1/m_b$, we estimate the two ratios ${\cal
B}^{(nl)}/{\cal B}^{(f+a)}$ and ${\cal B}^{(f+a)}/{\cal B}^{(f)}$
which can make sense about the involved physics. It is found that
for $\mu$ ranging from $\mu_b/2$ to $2\mu_b$ the effects of
annihilation topology can make the branching ratios ${\cal
B}^{(f)}(B \to K \pi)$ increase by $(20-30)\%$, whereas the soft
and hard effects modify ${\cal B}^{(f+a)}(B \to K \pi)$ at a level
of about $10\%$.

As emphasized, most of the inputs suffer from theoretical
uncertainties, which can affect the numerical results and should
be carefully examined. Obviously, these uncertainties are related
to the non-perturbative QCD, about which a solid knowledge is
absent at present. Thus we are not so ambitious to make a complete
quantitative estimate of them, instead, we just list the main
sources of uncertainties which we can conjecture, and discuss how
they influence the present results. By our observation, the most
important  sources of uncertainties are the following: (1)
\textit{Distribution amplitudes.} The models, which are here
adopted for the distribution amplitudes of light mesons, are based
on an expansion in conformal spin. For the leading twist-2
distribution amplitudes, the asymptotic forms are known precisely
and the non-asymptotic corrections, in spite of their
model-dependence, are also believed to be under control. In
contrast, little is known about the uncertainties in higher-twist
wavefunctions of light-mesons and distribution amplitudes of $B$
mesons. Therefore, this could have a significant impact on the
reliability of theoretical prediction. For example, the
higher-twist wavefunctions can  greatly affect the accuracy of the
$1/m_b$ correction parts and the hard spectator scattering
contributions, which are proportional to $1/\Lambda_{QCD}$. When
we carry out the calculation by using the extensively adopted form
of the B meson wavefunctions,  a considerable uncertainty may
exist. (2)\textit{"Chirally enhanced" trems}.  A term proportional
to the parameter $r_\chi^K$ appears in the calculations of the
matrix elements of the $(S-P)(S+P)$ operators. The part is called
the "chiral enhancement" term, being formally $1/m_b-$suppressed,
but numerically large. Because the factor is sensitive to the
current quark masses whose precise values are not known, their
variations may cause a considerable uncertainty in the numerical
results. (3) \textit{Form factors.} One may believe that
long-distance effects dominate the hadronic matrix elements of the
heavy-to-light transitions from a naive power counting. If it is
true, the LCSR results for the form factors, which are used in our
calculations, should be relatively reliable. Nevertheless, there
are other viewpoints contrary to it \cite{HN}, that is, the
short-distance contributions are predominant over long-distance
ones so that PQCD is applicable in this case. A better
understanding of the uncertainties due to the form factors asks
for a clarification of the transition mechanism. (4) \textit{Decay
constants.} In QCD sum rule calculations the decay constants for
the B mesons are sensitive to the b quark mass, as we know. In
contrast, the decay constants of the light pseudoscalar mesons
have been experimentally measured to high accuracy, the
uncertainties from this part is the least. (5) \textit {Final
state interactions(FSI).} Besides those effects which have already
been estimated, the FSI effects may play a non-negligible role in
the evaluation of decay amplitudes. These effects are usually
regarded as being power suppressed in $1/m_b$ owing to the large
energy released in the B decays. Unfortunately, so far, we lack a
convincing quantitative evaluation on them. The $B\to K\pi$ modes
can receive the final state rescattering contributions from a
number of intermediate states. The FSI contributions from the
$B\to \overline{D}_s D$ followed by $\overline{D}_s D\to K\pi$ are
anticipated to be especially important. The reason is obvious that
the $B\to \overline{D}_sD$ modes are $CKM$ favored and thus can
provide a large branching ratio; on the other hand, they are
typically dominated by long-distance dynamics and violate the QCD
factorization. (6)\textit {Other sources of power suppression.} In
SCET a new source of power suppressed contributions has been
identified in study on the color suppressed process $\bar{B^0}\to
D^0 \pi^0$. It is necessary to explore the corresponding impacts
on $B\to K\pi$. In addition, there might be other new sources of
power suppressed contributions awaiting to be dug out and their
effects should be estimated. (7)\textit {Theoretical approach.}
The numerical results presented here are based on a combination of
the QCD factorization formula, PQCD approach and LCSR method. A
possible inconsistency might manifest as the calculations are done
by using all the three different approaches. Of course, to avoid
this ambiguity is to do calculation within the same theoretical
framework. The LCSR approach provides such a possibility.
Practical manipulation, however, would be considerably difficult
in view of the existing complications in technique.

Once all these uncertain factors are taken into account and
clarified, either theoretically or phenomenologically, with our
new knowledge, the numerical results presented here can and should
eventually be updated.

\begin{center}
\bf{V. SUMMARY}
\end{center}

We obtain the LCSR results for the $1/m_b$ suppressed soft and
hard corrections to the $B\rightarrow K \pi$ decays. Then
combining them with the previously obtained results of QCD
factorization and estimates of annihilation topology, the
branching ratios are calculated and the numerical impacts of such
effects, as an order of magnitude estimate of the overall
power-suppressed effects, are investigated to subleading order in
$1/m_b$. Also, the important sources of uncertainties are
discussed in some details.

Subject to a quantitative estimate of the existing uncertainties,
for $\gamma=60^{\circ} \pm 20^{\circ}$ we have observed the
following. (1) The $1/m_b$ power suppressed effects resulting from
the annihilations and soft and hard exchanges can make the
branching ratios getting an increase of $(20-30)\%$ with respect
to the results of the QCD factorization, depending on concrete
decay modes. (2) The annihilation effect is predominate over the
soft and hard ones, which modify the branching ratios by only
$10\%$ of the results which include the ${\cal O }(\alpha_s)$
radiative corrections and annihilation effects.

Our present work helps to make sense about the contributions of
the soft and hard-gluon exchange along with other power suppressed
terms to the $B\rightarrow K\pi$ decays. No doubt, it still is too
early to draw a decisive conclusion at present whether or not the
theoretical estimates can accommodate the experimental data and
the power-suppressed soft and hard effects are negligibly small in
the $B\to K\pi$ decays. We have to await the improvement in
experiment and progress in theoretical or phenomenological study
on, amongst other things, the higher-twist wavefunctions as well
as behaviors of FSI effects in the heavy quark limit. Certainly, a
better understanding of the other sources of uncertainties is
indispensable to arrive at a reliable conclusion. Anyhow a further
investigation, whether theoretical or phenomenological, on the
power-suppressed effects in charmless B decays may be needed.

\begin{center}
\bf{ACKNOWLEDGEMENT}
\end{center}

This work is in part supported by the National Natural Science
Foundation of China.

%%%%%%%%%%%%%%%%%%%% Fig. 1%%%%%%%%%%%%%%%%
\newpage
\begin{figure}

\vspace{3.0cm}

\begin{center}

\includegraphics[scale=0.5]{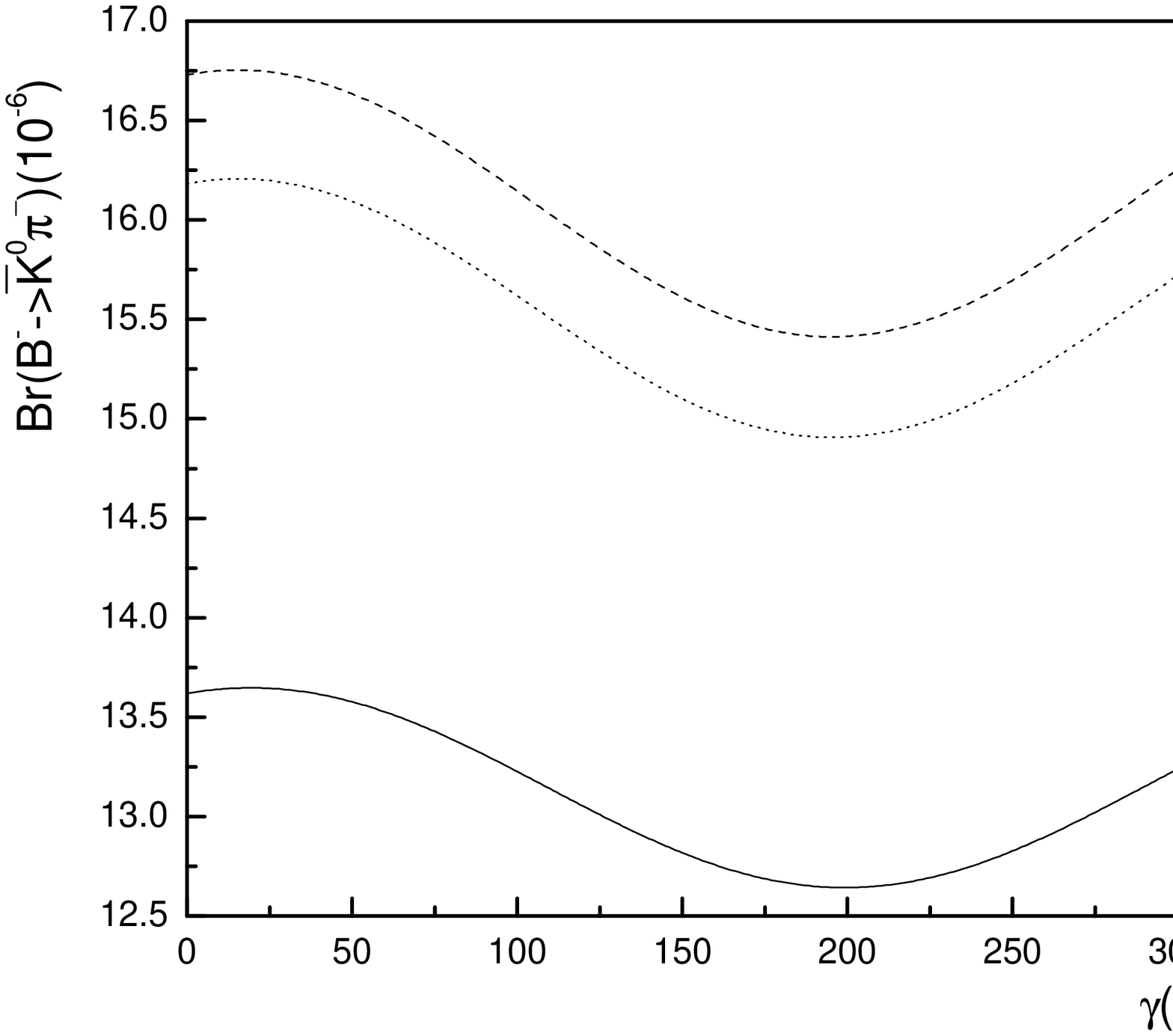}
\vspace{-5.cm} \caption{${\cal B}(B^{-}\to \bar{K}^0 \pi^{-})$
versus the weak phase $\gamma$. The solid, dotted and dashed lines
correspond to ${\cal B}^{(f)}(B^{-}\to \bar{K}^0 \pi^{-})$, ${\cal
B}^{(f+a)}(B^{-}\to \bar{K}^0 \pi^{-})$ and ${\cal
B}^{(nl)}(B^{-}\to \bar{K}^0 \pi^{-})$, respectively.}
\end{center}
\label{block1}
\end{figure}
\newpage

\begin{figure}[htbp]

\vspace{1.5cm}

\begin{center}

\includegraphics[scale=0.5]{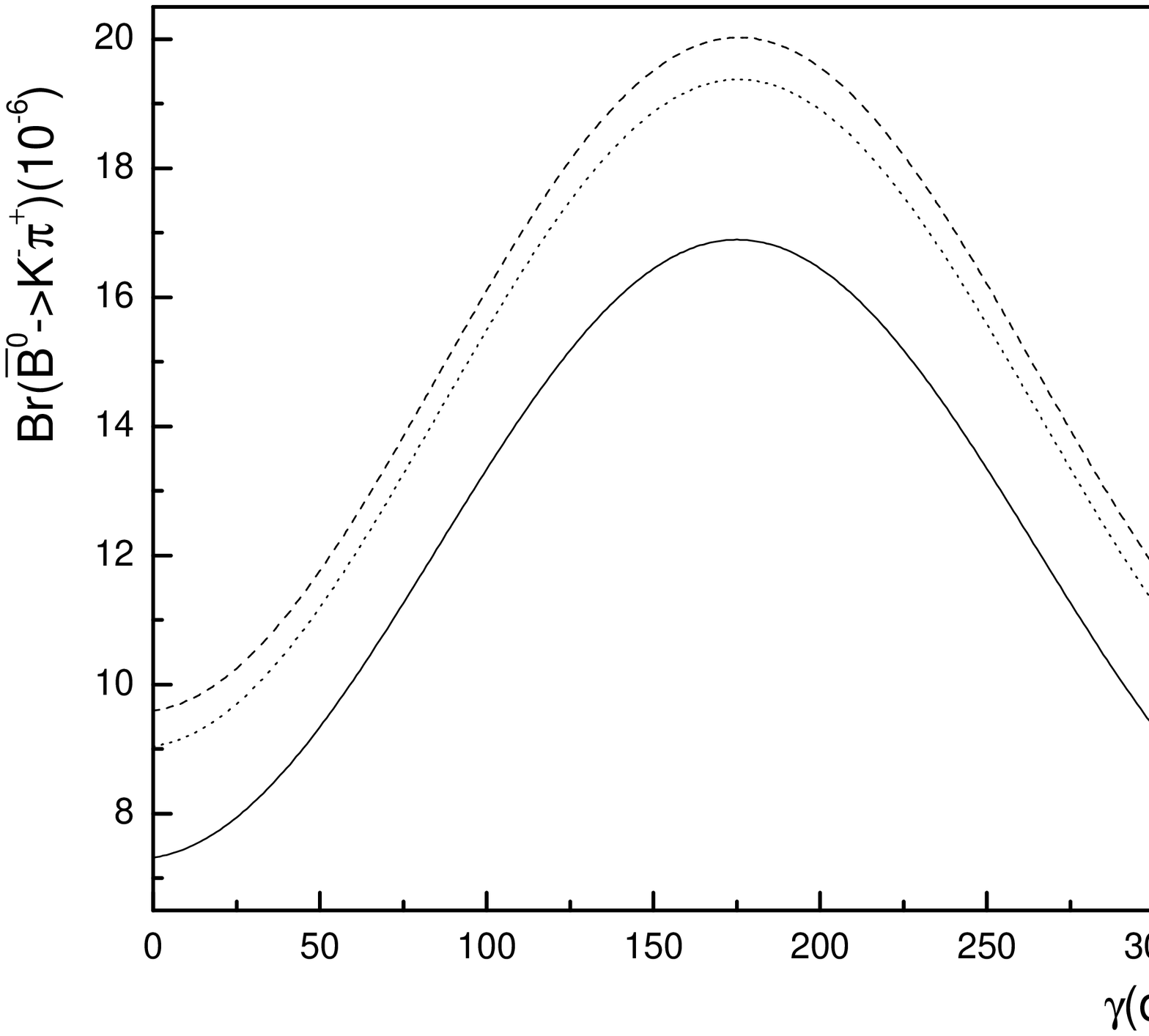}

\vspace{-5cm}

\caption{${\cal B}(\bar{B}^0\rightarrow K^-\pi^{+})$ versus the
weak phase $\gamma$. The definitions of the lines are the same as
in Figure 1.}
\label{block1}
\end{center}
\end{figure}
%%%%%%%%%%%%%%%%%%%%%%%%%%%%%%%%%%%%%%%%%%%

%%%%%%%%%%%%%%%%%%%% Fig. 3%%%%%%%%%%%%%%%%
\newpage
\begin{figure}[htbp]

\vspace{3.0cm}

\begin{center}
\includegraphics[scale=0.5]{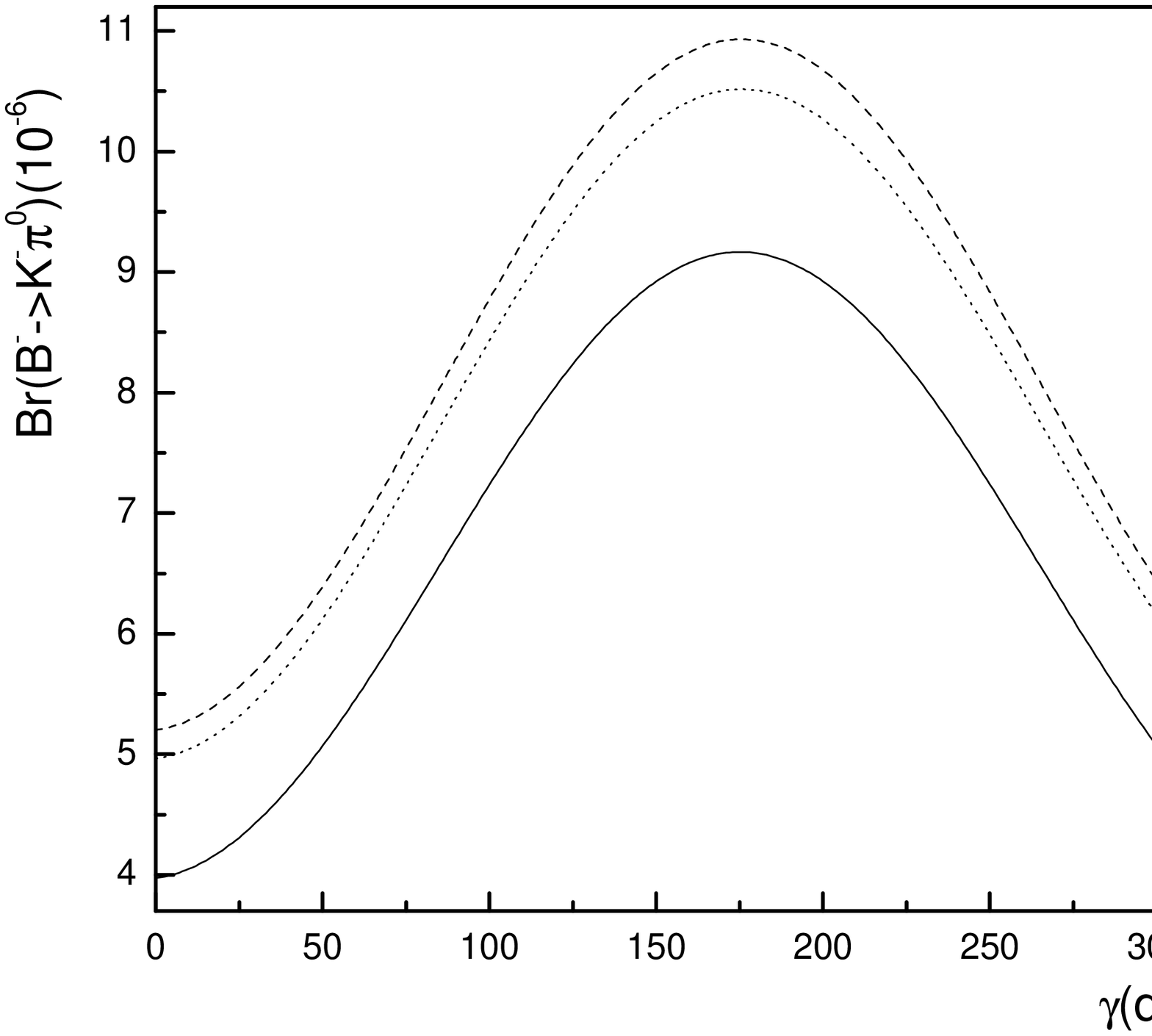}
\vspace{-5.cm}
 \caption{${\cal B}(B^-\rightarrow K^-\pi^{0})$ versus the weak
phase $\gamma$. The definitions of the lines are the same as in
Figure 1.}
 \label{block1}
\end{center}
\end{figure}
%%%%%%%%%%%%%%%%%%%%%%%%%%%%%%%%%%%%%%%%%%%
%%%%%%%%%%%%%%%%%%% Fig. 4%%%%%%%%%%%%%%%%
%\newpage
\begin{figure}[htbp] \vspace{1.0cm}
\begin{center}
\includegraphics[scale=0.5]{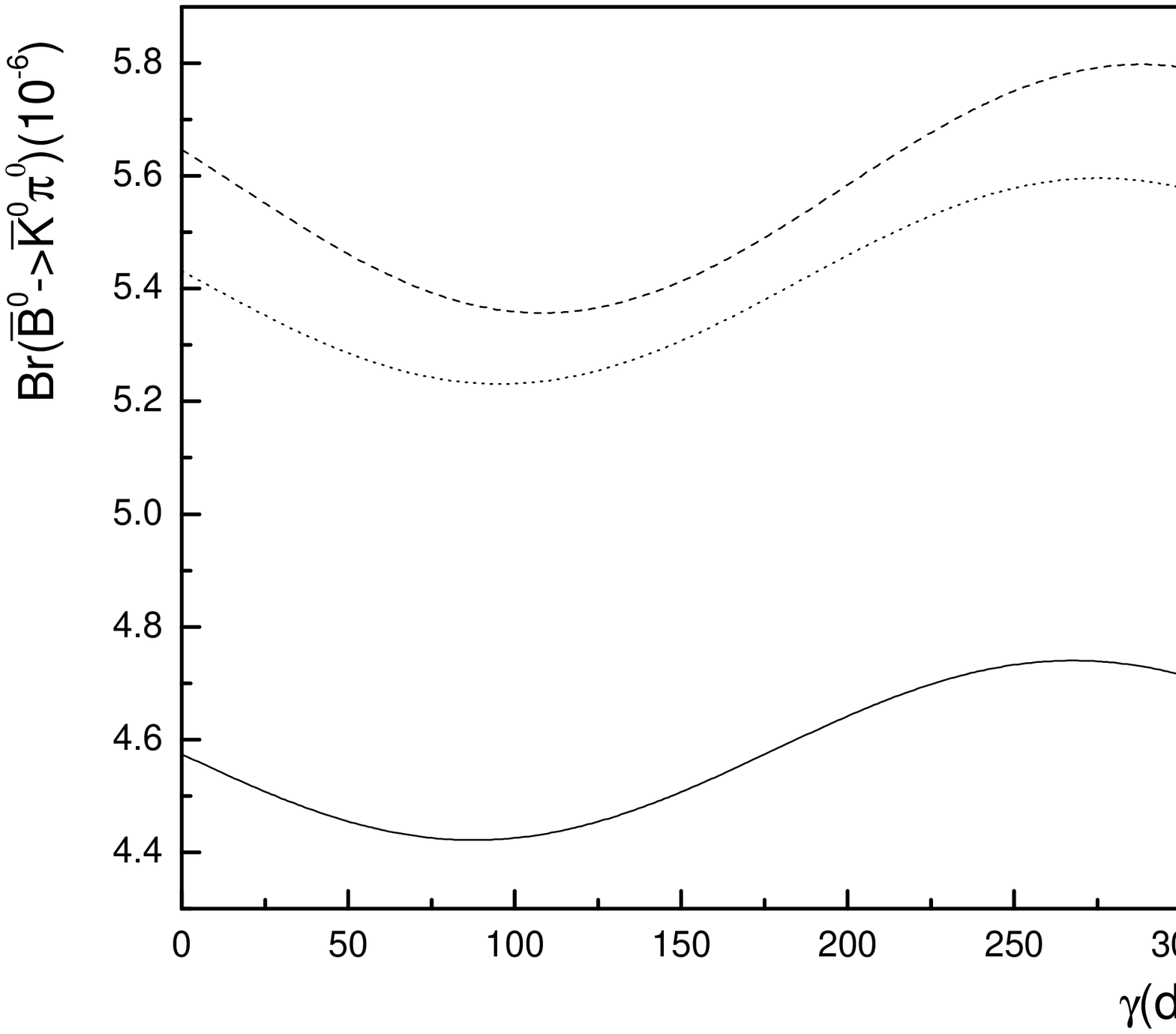}
\vspace{-5.cm} \caption{${\cal B}(\bar{B}^0\rightarrow
\bar{K}^0\pi^{0})$ versus the weak phase $\gamma$. The definitions
of the lines are the same as in Figure 1.} \label{block1}
\end{center}
\end{figure}
%%%%%%%%%%%%%%%%%%%%%%%%%%%%%%%%%%%%%%%%%%%

\begin{figure}[htbp] \vspace{3.0cm}
\begin{center}
\includegraphics[scale=0.5]{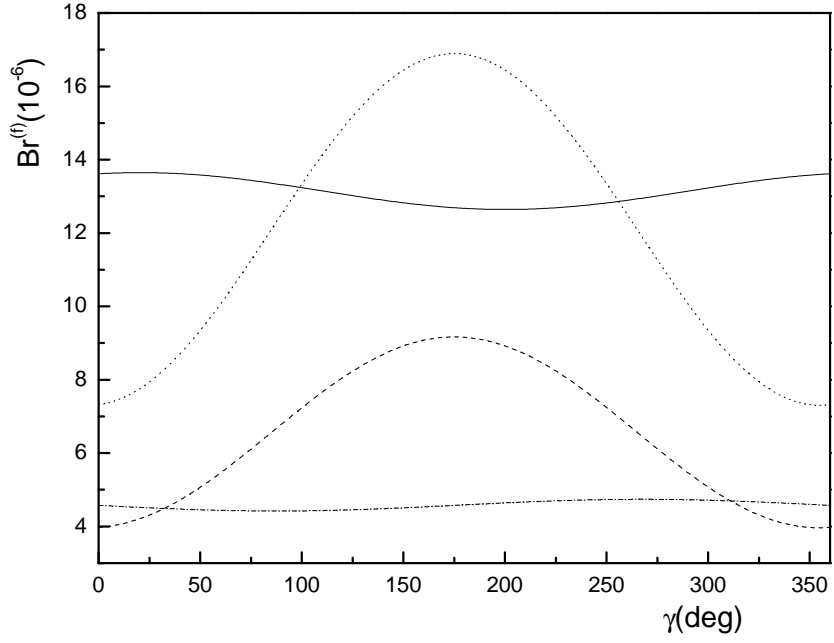}
\vspace{-5.cm} \caption{${\cal B}^{(f)}(B\to K\pi)$ versus the
weak phase $\gamma$. The solid, dotted, dashed and dash-dotted
lines correspond to $B^{-}\to \bar{K}^0 \pi^{-}$,
$\bar{B}^0\rightarrow K^-\pi^{+}$, $B^-\rightarrow K^-\pi^{0}$ and
$\bar{B}^0\rightarrow \bar{K}^0\pi^{0}$, respectively.}
\label{block1}
\end{center}
\end{figure}
%%%%%%%%%%%%%%%%%%%%%%%%%%%%%%%%%%%%%%%%%%%
%%%%%%%%%%%%%%%%%%%% Fig. 6%%%%%%%%%%%%%%%%

\begin{figure}[htbp] \vspace{1.0cm}
\begin{center}
\includegraphics[scale=0.5]{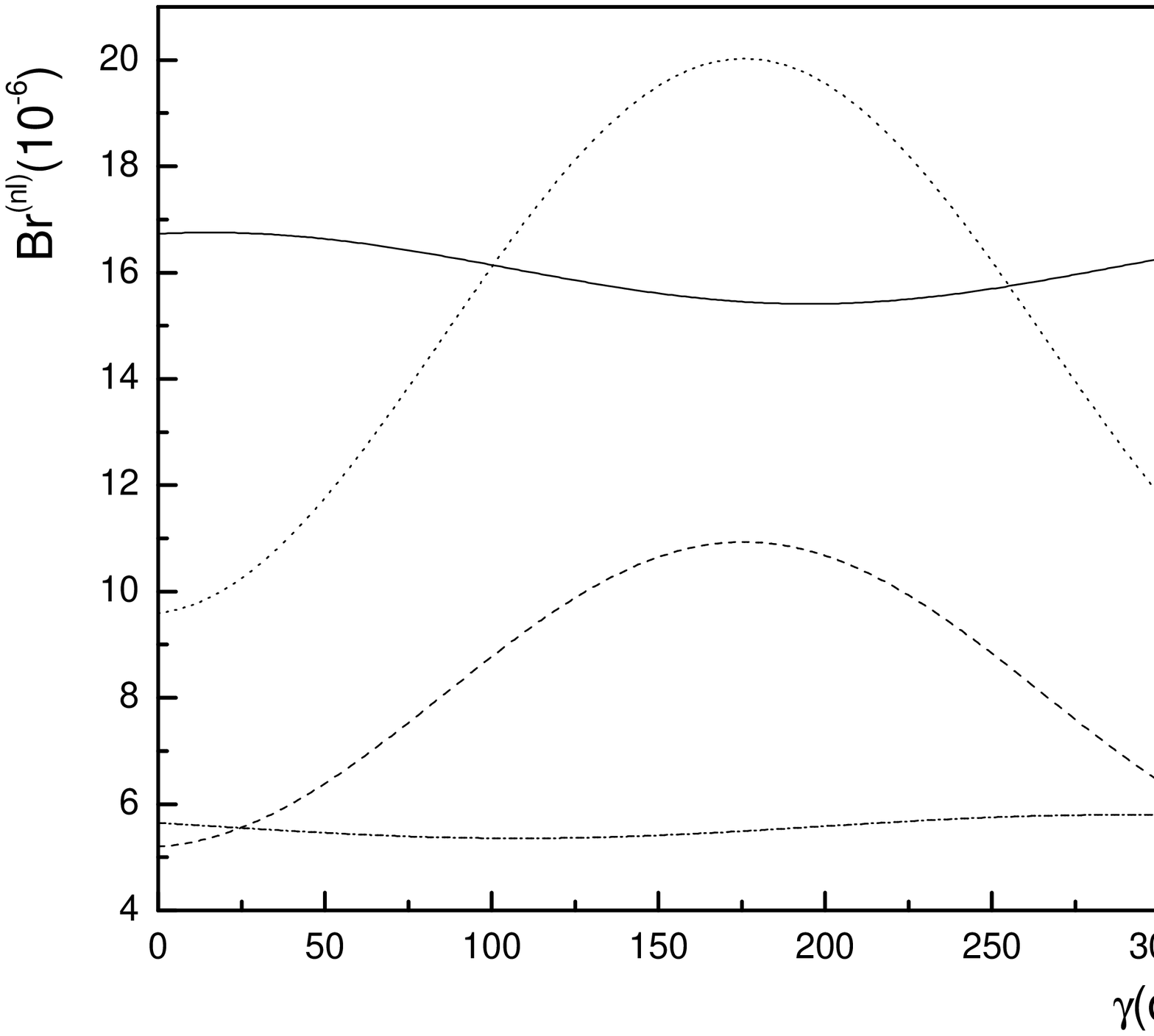}
\vspace{-5.cm} \caption{${\cal B}^{(nl)}(B\to K\pi)$ versus the
weak phase $\gamma$. The definitions of the lines are the same as
in Figure 5.} \label{block1}
\end{center}
\end{figure}
%%%%%%%%%%%%%%%%%%%%%%%%%%%%%%%%%%%%%%%%%%%

\end{document}